\documentclass[
    ,final            % use final for the camera ready runs
  ]
  {aipproc}

\layoutstyle{6x9}

\newcommand{\beq}{\begin{equation}}
\newcommand{\eeq}{\end{equation}}
\newcommand{\ba}{\begin{array}{ccc}}
\newcommand{\ea}{\end{array}}
\newcommand{\nn}{\nonumber}
 
\def\beqn{\begin{eqnarray}}
\def\eeqn{\end{eqnarray}}

\def\<{\langle}
\def\>{\rangle}

\def\cond{ \frac{b g^2}{32 \pi^2 } G_{\mu \nu}^a G^{\mu \nu a}}

\usepackage{graphicx}
\usepackage{amsmath}
\usepackage{amssymb}
\usepackage{psfrag}
\usepackage{epsfig}

\begin{document}

\title{ Vacuum Energy,  EoS, and the Gluon Condensate at Finite Baryon Density in QCD}
  
%{gluon condensate, equation of states, color superconductivity}

 \author{ Ariel~R.~Zhitnitsky}{ address={ Department of Physics and
Astronomy, University of British Columbia, Vancouver, BC, Canada,V6T 1Z1}}

\begin{abstract}
The Equation of States (EoS) plays the crucial role in all studies of neutron star properties.
 Still, a microscopical understanding   of EoS remains  largely an unresolved problem.
 We use 2-color QCD as a model to study the dependence of vacuum energy
(gluon condensate in QCD) as function of chemical potential $\mu\ll \Lambda_{QCD}$
where we find very strong and unexpected dependence on $\mu$.
We present the arguments suggesting that similar behavior  may occur 
  in 3-color QCD in the color superconducting phases.
Such a study may be of importance for analysis    of EoS
when phenomenologically relevant parameters   (within such models as MIT Bag model or NJL model)
are fixed at zero density while the   region of study lies at much higher densities
not available for terrestrial tests. 

\end{abstract}

\maketitle
  
\section{Introduction}

This talk is based on few recent   publications  with Max Metlitski \cite{Metlitski:2005db}.
Neutron stars represent one of the densest concentrations of matter in our universe. 
 The properties of super dense matter are fundamental to our understanding of nature   
  of nuclear forces as well as the underlying theory of strong interactions, QCD.
  Unfortunately, at present time,  we are not in a position to    answer many 
  important questions starting from fundamental   QCD lagrangian.
  Instead, this problem is usually attacked by using some  phenomenological models such 
   as MIT Bag model or NJL model. Dimensional parameters (e.g. the vacuum energy) for these models
   are typically fixed by using available experimental data at zero baryon density.   
   Once the  parameters are fixed, the analysis of EoS or other quantities    is typically performed 
  by    assuming that the
  parameters of the models (e.g. bag constant) at nonzero $\mu$ are the same as  at $\mu=0$. 
  
  The main lesson to be learned 
  from the calculations presented below can be formulated as follows: the standard assumption (fixing the parameters of a model  at $\mu=0$ while calculating the observables at    nonzero $\mu$)
   may be badly violated in QCD.
  
   The problem of density dependence of the 
   chiral and gluon condensates in QCD has been addressed
   long ago in\cite{Cohen:1991nk}. The main  motivation of ref.\cite{Cohen:1991nk}
   was the application of  the QCD sum rules technique to study some hadronic properties in the nuclear matter environment.
   The main result of that studies is--  the effect is small. More precisely, 
   at nuclear matter saturation density  
    the change of the gluon condensate is  only about $5\%$.  Indeed,
     in the chiral limit the variation of the gluon condensate with density  can  be expressed  
    as follows\cite{Cohen:1991nk},
\beq
\label{cohen}
\langle \cond\rangle_{\rho_B} -
\langle\cond\rangle_{0}  
  = -m_N\rho_B +0(\rho^2_B), ~~ b=\frac{11 N_c -2 N_f}{3},
\eeq
where the standard expression for the conformal anomaly
is used, $\Theta^{\mu}_{\mu} = - \frac{b g^2}{32 \pi^2}
G^a_{\mu\nu} G^{a\mu \nu}$.  We should note here that the variation 
of the gluon condensate is well defined observable (in contrast with the gluon condensate itself)
because 
the perturbative (divergent)  contribution cancels in eq.(\ref{cohen}).
  The most important consequences of this formula: a) the variation 
of the gluon condensate   is small numerically, and b) the absolute value of the condensate
decreases when the baryon density increases. Such a behavior    can be interpreted as 
 due to the    suppression of the non-perturbative  QCD fluctuations with increase of the  baryon density.
    
  Our ultimate goal here is to understand the behavior of the vacuum energy  
  (gluon condensate ) as a function of $\mu$ for color superconducting (CS) phases\cite{ARF},
\cite{Rapp:1998zu}. It is clear that the problem in this case is drastically different from nuclear matter
analysis \cite{Cohen:1991nk}
 because  the system becomes relativistic and binding energy ($\sim \Delta$)
 per baryon charge is order of $\Lambda_{QCD}$ in contrast 
 with   $\leq 2\%$ of the nucleon mass at  nuclear saturation density. 
 The quark-quark interaction  also becomes essential in CS phases such that the
small density  expansion (valid for dilute noninteracting nuclear matter)   used to derive (\ref{cohen}) can  not  be justified any more.

 Unfortunately, we can not  answer the questions  on $\mu$ dependence
of the vacuum energy  in 
 real   $QCD(N_c=3)$. However, these questions  can be formulated and can be answered in more simple model $ QCD(N_c=2)$ due to the  extended symmetry of this model.
 Some lessons for the real life with $N_c=3$ can be learned from our analysis, see below.
 \section{Gluon Condensate for $ QCD(N_c=2)$ }
We start from the equation for the conformal anomaly, \beq
\label{Tmm} \Theta^{\mu}_{\mu} = - \frac{b g^2}{32 \pi^2}
G^a_{\mu\nu} G^{a\mu \nu} + \bar{\psi} M \psi , ~~~~~~~ b = \frac{11}{3} N_c - \frac{2}{3} N_f = 6
\eeq  
   For massless quarks and in the absence of chemical
potential, eq.\,(\ref{Tmm}) implies that the QCD vacuum carries a
negative non-perturbative vacuum energy due to the gluon
condensate.

Now, we can use the effective Lagrangian \cite{kogut2}
\beq
\label{Lch}
 {\cal L} = \frac{F^2}{2} Tr \nabla_{\nu} \Sigma
\nabla_{\nu} \Sigma^{\dagger} ,~~ 
  \nabla_0 \Sigma = \partial_0 \Sigma -
\mu\left[ B  \Sigma +  \Sigma  B^T\right] 
\eeq
to calculate
the change in the trace of the energy-momentum tensor $\langle
\theta^{\mu}_{\mu}\rangle$ due to a finite chemical potential
$\mu \ll \Lambda_{QCD}$. The energy density $\epsilon$ and
pressure $p$ are obtained from the free
energy density ${\cal F}$, 
\beqn \epsilon = {\cal F} + \mu n_B,~~~
p = -{\cal F}. 
\eeqn 
Therefore, the conformal anomaly implies,
\beqn \label{DG2}\langle\cond\rangle_{\mu, m} -
\langle\cond\rangle_0 = \nn\\
-4 \left({\cal F}(\mu, m) - {\cal
F}_0\right)- \mu n_B(\mu, m) + \langle \bar{\psi} M \psi
\rangle_{\mu, m},
\eeqn 
where the subscript $0$ on an
expectation value means that it is evaluated at $\mu = m = 0$. 
 Now we notice that all quantities on the right hand side are known from the previous calculations \cite{kogut2}, therefore the variation of $\langle G^2_{\mu\nu}\rangle$ with $\mu$ can be explicitly calculated.
 As
expected, $\langle G^2_{\mu\nu}\rangle$ does not depend on $\mu$
in the normal phase $\mu < m_{\pi} $ while in  the superfluid  phase $\mu > m_{\pi} $   
this dependence can be represented as follows \cite{Metlitski:2005db},
 \beq
\label{G2M} \langle\cond\rangle_{\mu, m} -
\langle\cond\rangle_{\mu=0, m} = 4 F^2 (\mu^2
-m^2_{\pi}) \left(1 - 2
\frac{m^2_{\pi}}{\mu^2}\right).\eeq 
  The behavior of the condensate is quite interesting: it decreases
with $\mu$ for $m_\pi < \mu < 2^{1/4} m_\pi$ and increases
afterwards. The qualitative difference in the behaviour of the
gluon condensate for $\mu \approx m_\pi$ and for $m_\pi \ll
\mu \ll \Lambda_{QCD}$ can be explained as follows. Right after
the normal to superfluid  phase transition occurs, the baryon density
$n_B$ is small and our system can be understood as a weakly
interacting gas of diquarks. The pressure of such a gas is
negligible compared to the energy density, which comes mostly from
diquark rest mass. Thus, $\langle\Theta^{\mu}_{\mu}\rangle$
increases with $n_B$   in 
 precise correspondence  with  the ``dilute" nuclear matter  case (\ref{cohen}).
  On the other hand, for $\mu \gg
m_{\pi}$, energy density is approximately equal to pressure, and
both are mostly due to self-interactions of the diquark
condensate. Luckily, the effective Chiral Lagrangian (\ref{Lch})
gives us control over these self-interactions as long as $\mu
\ll \Lambda_{QCD}$.  

The main lesson to be learned for real $ QCD (N_c=3) $ from exact results  
discussed above is as follows. The transition to the CS phases
is expected to occur\footnote{  $\mu$ here for $ QCD (N_c=3) $ is normalized as the quark
 (rather than the baryon) chemical potential}  at $\mu_c\simeq 2.3\cdot  \Lambda_{QCD}$\cite{Toublan:2005tn},\cite{Zhitnitsky:2006sr}
in contrast with $\mu_c=m_{\pi}$ for transition to superfluid phase for $N_c=2$ case. The binding energy, the gap, the quasi -particle  
masses are also 
expected to be the same order of magnitude $\sim \mu_c$.
 This is in drastic contrast with nuclear matter case
 when binding energy  
is very small.
 At the same time, $ QCD (N_c=2) $ represents a nice model where 
the  binding energy, the gap, the masses of quasi -particles  
  carrying the baryon charge are the same order of magnitude.   This model explicitly shows that
the gluon condensate  can experience extremely nontrivial behavior as function of $\mu$.
We expect a similar behavior for $ QCD (N_c=3) $ in CS phases when 
function of $ {m^2_{\pi}}/{\mu^2}$ in (\ref{G2M}) is replaced by some  function of 
$ {\mu_c}/{\mu^2} $ for $N_c=3$.
 We should note in conclusion that the recent lattice calculations  \cite{Hands:2006ve},\cite{Alles:2006ea}
  are consistent with our prediction (\ref{G2M}).

\end{document}